# Edge-modulated perpendicular magnetic anisotropy in [Co/Pd]$_n$ and $L1_0$-FePt thin film wires

Jinshuo Zhang[1], Pin Ho[1], Jean Anne Currivan-Incorvia[2,3], Saima A. Siddiqui[3], M. Baldo[3], Caroline A. Ross[1]

[1]*Department of Materials Science and Engineering, Massachusetts Institute of Technology, Cambridge, MA 02139 USA*

[2]*Department of Physics, Harvard University, Cambridge, MA 02138 USA*

[3]*Department of Electrical Engineering and Computer Science, Massachusetts Institute of Technology, Cambridge, MA 02139 USA*

**Abstract:**

Thickness modulation at the edges of nanostructured magnetic thin films is shown to have important effects on their perpendicular magnetic anisotropy. Thin film wires with tapered edges were made from [Co/Pd]$_{20}$ multilayers or $L1_0$-FePt films using liftoff with a double-layer resist. The effect of edge taper on the reversal process was studied using magnetic force microscopy (MFM) and micromagnetic modeling. In [Co/Pd]$_{20}$ the anisotropy was lower in the tapered edge regions which switched at a lower reverse field compared to the center of the wire. The $L1_0$-FePt wires showed opposite behavior with the tapered regions exhibiting higher anisotropy.

Thin film materials with strong perpendicular magnetic anisotropy (PMA) such as [Co/Pd]$_n$ multilayers and $L1_0$-FePt are useful for a variety of magnetic recording and logic devices because their high anisotropy promotes thermal stability at low dimensions [1-5]. The PMA in [Co/Pd]$_n$ multilayers originates from interfacial anisotropy which is induced by lattice mismatch strain [6-8]. A greater PMA is achieved by increasing the number of interfaces or optimizing the interfacial structure in [Co/Pd]$_n$ multilayers [1, 9]. In contrast, the high PMA in a (001)-oriented $L1_0$-FePt film, which can be described as alternating monolayers of Fe and Pt atoms, originates from the spin orbit coupling of Pt and strong hybridization between the Pt $5d$ and Fe $3d$ electronic states. (001)-oriented $L1_0$-FePt films with high PMA are made by depositing FePt on top of single crystalline substrates with (001) orientation, such as MgO, $KTaO_3$ and $SrTiO_3$. Heat treatment during a post-deposition anneal (PA) or rapid thermal anneal (RTA) introduces tensile strain in the FePt film, which favors formation of ordered $L1_0$-FePt with (001) texture [10-12].

The study of the magnetic reversal processes in thin film materials with high PMA is of significant importance [13-15], especially in patterned nanoscale structures since they play a key role in many device applications. The reversal is affected by the magnitude of PMA as well as the shape and size of the nanostructure which affect the micromagnetic configuration and the nucleation and propagation of domains. The nature and geometry of the edges is particularly important because of demagnetizing fields which affect dynamic magnetic properties such as edge modes and domain wall motion [16, 17], and because edge roughness can provide nucleation or pinning sites for domain walls. Magnetic properties including PMA and magnetization also vary near the edges because the edges may consist thinner layers or of more disordered material, depending on the fabrication process. Thus, control of the edges of nanostructures provides a path to further design the magnetic properties. In this article, we describe how edge tapering affects PMA and switching of nanostructures made from both $[Co/Pd]_n$ multilayers and $L1_0$-FePt films, allowing PMA to be enhanced or reduced near the edge, and show how this affects domain nucleation.

The samples with edge taper are produced using methyl methacrylate (MMA) - polymethyl methacrylate (PMMA) double-layer resist, where each resist layer has different solubility in the same developer after exposure [18]. Electron-beam lithography is used to expose the patterns in the bilayer. After exposure, the bottom layer resist dissolves faster than the top layer in the developer, resulting in an undercut profile. When used in a liftoff process with sputtering or other physical vapor deposition processes, this produces features with a tapered profile [19] and lower sidewall roughness than that obtained from single-layer resist, as shown in Fig. 1(a). This process was applied to [Co/Pd]$_n$ multilayers grown on Si/SiO$_2$ substrates and $L1_0$-FePt grown on MgO (001) substrates, which reduces the PMA in [Co/Pd]$_n$ and enhances the PMA in $L1_0$-FePt at the edges.

To form the double-layer resist, copolymer MMA ( 8.5 kg/mol) 3% solution in anisole was spin-coated at 4 krpm for 60 s on Si/SiO$_2$ or MgO substrate for [Co/Pd]$_n$ and FePt samples, respectively and baked at 150 °C for 90 s to produce 30 nm-thick film. PMMA (950 kg/mol) 3% solution in anisole was then spin-coated at 4 krpm for 60 s on top and baked at 180 °C for 90 s to produce an 50 nm thick overlayer. A Raith 150 electron beam lithography tool was used for exposure with 10 kV acceleration voltage and 130 µC cm$^{-2}$ dosage. The exposed samples were then developed using methyl isobutyl ketone : isopropyl alcohol = 1 : 3 solution for 2 minutes at 21 °C, rinsed with isopropyl alcohol and dried by nitrogen blow gun. A cross-section SEM of the resulting resist pattern is shown in Fig. 1(c), where the undercut is about 150 nm. For comparison, single-layer PMMA resist patterns were also made with the same coating method, exposure and development conditions.

Magnetic films were deposited onto both resist patterns, then lifted off by placing the samples in n-methyl-2-pyrrolidone at 90 °C for 2 hours. Cross-sectional AFM profiles of exemplary nanowires made using single-layer and double-layer resist processes are shown in Fig. 1(b), illustrating a tapered edge with reduced crowning using double-layer resist. Fig. 1(d) shows the actual wire width measured from AFM vs. nominal wire width according to lithography design for [Co/Pd]$_n$ multilayers. Wires made using double-layer resist had a tapered shape with top width 10-25 % higher than nominal and base width about twice as large. For example, a nominally 300 nm wide wire had a base width of 780 nm and a top width of 370 nm. For the same nominal width, the fabricated width of nanowires from single-layer resist was intermediate between the bottom and top widths of wires from double layer resist. The higher-than-nominal width was due to over-exposure in order to optimize the resist profile.

The [Co (0.6 nm)/Pd (1.2 nm)]$_n$ multilayers were prepared by UHV magnetron sputter deposition of Co and Pd layers alternately under an Ar pressure of 10 mTorr and a base pressure better than $3 \times 10^{-7}$ Torr. A power of 25 W was used, resulting in 0.62 and 1.86 nm min$^{-1}$ deposition rate for Co and Pd, respectively. The Co and Pd layer thickness were chosen to optimize the PMA in a continuous film. Films with different layer repetitions of $n = 10, 15, 20, 25$ were prepared and out-of-plane magnetic hysteresis loops were measured using VSM. The coercive field increased from $1020 \pm 50$ Oe for $n = 10$ to $2250 \pm 50$ Oe for $n = 15$, shown in Fig. 2(a). Further increasing the repetitions up to 25 only increased the coercive field slightly to $2300 \pm 50$ Oe. Saturation magnetization M$_s$ of the four samples was between 200 and 230 emu cm$^{-3}$ and there was no systematic variation with *n*. The uniaxial anisotropy of the material, defined by $K_u = \frac{1}{2} M_s H_k$ with anisotropy field H$_k$ determined from extrapolation of the in-plane loops to saturation, was $5.90 \pm 0.50 \times 10^5$ erg cm$^{-3}$, $7.45 \pm 0.60 \times 10^5$ erg cm$^{-3}$, $7.78 \pm 0.60 \times 10^5$ erg cm$^{-3}$ and $8.20 \pm 0.60 \times 10^5$ erg cm$^{-3}$ for *n* = 10, 15, 20, and 25, respectively. This is the net anisotropy which is the sum of the shape anisotropy and the interfacial anisotropy. Calculation of the anisotropy as the difference between the areas of the hard axis loop and the anhysteretic easy axis loop gave similar results.

[Co (0.6 nm)/Pd (1.2 nm)]$_{20}$ multilayers with a coercive field of $2280 \pm 50$ Oe and thickness 36 nm were used in further experiments. Nanowires with nominal widths of 200 nm, 300 nm, 500 nm and 800 nm using either double or single-layer resist liftoff process were made. The measured wire widths are shown in Fig. 1(d).

Magnetic reversal of patterned [Co (0.6 nm)/Pd (1.2 nm)]$_{20}$ was studied using MFM. A +10 kOe field in the *+z* direction was first applied to fully saturate the magnetization out-of-plane, then a smaller reverse field $H_{REV}$ was applied in the –z direction. At $H_{REV} = -1$ kOe, reverse domains nucleated first at the tapered edge of the wires [Fig. 3(a)]. As shown in Fig. 3(b), when $H_{REV}$ increased to -2 kOe the reverse domains at the edge propagated towards the center, for nominal widths of 200 nm, 300 nm and 500 nm. For 800 nm wide wires, reverse domains additionally nucleated at the center, characteristic of the unpatterned film. Increasing $H_{REV}$ to -3 kOe, above the coercive field, resulted in a nearly complete reversal of the tapered wires. In comparison, in the non-tapered wires, magnetization reversal occurred at $H_{REV} = -2$ kOe throughout the wire as shown in Fig. S1 of the supplemental materials [20].

The reduction of PMA at the edge of the tapered wires occurs because tapering leads to thinner Co and Pd layers in the multilayer which lowers $K_u$ [1]. In unpatterned films, reducing the Co layer thickness from 0.6 nm to 0.3 nm or reducing the Pd layer thickness from 1.2 nm to 0.5 nm eliminated PMA. As the wire thickness continues to decrease towards the edge, the Co and Pd layers become discontinuous which further degrades the interface-induced PMA. There may also be change in stoichiometry [21] if the arriving fluxes of Co and Pd have different angular distributions.

We now discuss the behavior of tapered PMA $L1_0$-FePt wires on MgO (001) substrate. The easy axis is along the c-axis of the tetragonal unit cell which is oriented out-of-plane. The PMA in $L1_0$-FePt is dependent on the strain at the interface with the MgO substrate and experimentally, the PMA decreases with increasing FePt thickness. Fig. 2(d) shows the XRD data of continuous FePt films with 10 nm and 20 nm thickness after a post-deposition vacuum anneal at 700 °C for 2 hours. The (001) family of peaks of FePt was observed in both films, confirming the existence of the ordered $L1_0$ phase. The ratio of superlattice peak to the fundamental peak intensity was $I_{(001)}/I_{(002)} = 0.57$ for the 10 nm film but only 0.49 for the 20 nm film, confirming better ordering for the 10 nm film [12]. The 20 nm film is also believed to include a larger proportion of (111)-oriented grains whose easy axis is tilted 36° with respect to the film plane. The magnetic moment contributed by the diamagnetic MgO substrate was measured in the in-plane and out-of-plane orientations and the VSM loops for FePt films were corrected by subtraction of the appropriate substrate contribution. The anisotropy was calculated from the VSM hysteresis loops in Fig. 2(c) in the same way as used for the Co/Pd films. Using $M_s = 700 \pm 50$ emu cm$^{-3}$ and $H_k = 76.5 \pm 0.1$ kOe for the 10 nm film and $64.6 \pm 0.1$ kOe for the 20 nm film, obtained by extrapolating the in-plane loops to saturation, we obtain $K_u^{10\ nm} = (2.68 \pm 0.20) \times 10^7$ erg cm$^{-3}$ and $K_u^{20\ nm} = (2.24 \pm 0.18) \times 10^7$ erg cm$^{-3}$. Therefore, the 10 nm film had better order and crystallographic texture, and higher anisotropy than the 20 nm film. This is expected to enable nucleation of reverse domains at a lower applied field in the thicker film. However, VSM also showed that the coercive field for the annealed continuous films is $H_c^{10\ nm} = 1700 \pm 100$ Oe and $H_c^{20\ nm} = 3300 \pm 100$ Oe. The higher coercivity for the 20 nm film, seen in both in-plane and out-of-plane loops, is attributed to a greater film inhomogeneity and pinning of domain wall motion as a result of the strain gradient and the presence of (111) oriented grains [22-24].

These differences suggest that the tapered edges of the $L1_0$-FePt would have a higher PMA than the wire center. The wires were 25 nm thick and prepared under the same conditions as the continuous films and annealed after liftoff. As shown in Fig. 4, magnetization reversal began in the tapered wires at $H_{REV} = -2$ kOe and the density of reverse domains was higher at the wire center than at the tapered edge. This is consistent with our expectation that lower PMA at the wire center facilitates DW nucleation for $L1_0$-FePt. Although preferential nucleation of reverse domains was observed at the wire center, the effect is less distinct compared to [Co/Pd]$_{20}$: despite the difference in domain density and size, the reversal process at the center and the edge both took place at $H_{REV} = -2$ kOe within the measurement resolution. When the reversal field increased to $H_{REV} = -3$ kOe, the proportion of reversed magnetic domains also increased and domains were present uniformly across the wire.

Micromagnetic simulations in [Co/Pd]$_n$ and $L1_0$-FePt nanowires were done using *OOMMF* [25]. The wire length was 3 $\mu$m and the cell size was $5 \times 5 \times 6$ nm$^3$ for [Co/Pd]$_{20}$ and $5 \times 5 \times 4$ nm$^3$ for FePt. The systems were 6-layers of cells thick, making the wire thickness in simulations (36 nm for [Co/Pd]$_{20}$ and 24nm for FePt) approximately the same as experiments. The edge taper was simulated by assigning 400 nm, 600 nm and 800 nm width to the top two, middle two and bottom two layers of cells, respectively. For [Co/Pd]$_{20}$, $M_s$ was 200 emu cm$^{-3}$, exchange constant was $A = 1.0 \times 10^{-6}$ erg/cm and uniaxial out-of-plane anisotropy was $K_u = 7.5 \times 10^5, 3.75 \times 10^5$ and 0 erg cm$^{-3}$ in the 36 nm, 24 nm and 12 nm thick regions, respectively. For $L1_0$-FePt wires, $M_s = 7.0 \times 10^2$ emu cm$^{-3}$, $A = 1.0 \times 10^{-6}$ erg/cm and $K_u = 2.0 \times 10^7, 2.2 \times 10^7$ and $2.4 \times 10^7$ erg cm$^{-3}$ in the 24 nm, 16 nm and 8 nm thick regions, respectively. Damping constant was set to 0.5 for rapid convergence. The initial magnetization was saturated in the $+z$ out-of-plane direction, then different fields in the $-z$ direction were applied and the magnetization state was calculated. The model had periodic boundary conditions along the wire length.

Fig. 5(a) shows the [Co/Pd]$_{20}$ model results. The 12 nm thick regions at the edge had an in-plane remanent state because the PMA was zero and the magnetization tilted out-of-plane as the thickness increased. The 12 nm thick edge tilted out-of-plane in a reverse field $H_z = -1$ kOe. On increasing the reverse field to $H_z = -1.4$ kOe, the 24 nm thick region was also reversed as in the third panel of Fig. 5(a). All the regions including the central 36 nm thick region were reversed at $H_z = -2$ kOe as in the last panel. After removing the $-1$ kOe or $-1.4$ kOe reverse field, the model relaxed to a remanent state similar to the first panel of Fig. 5(a), without residual domains. Such simulation results indicate that the lower PMA edge reverses before the higher PMA center similar to experiment results. The simulation did not reproduce domain formation and propagation, presumably because the model did not include local anisotropy variations due to, for example, polycrystallinity in the film or shape variation in the resist mask, nor thermal fluctuations.

In comparison, as shown in Fig 5(b), in simulations for $L1_0$-FePt thin film wires, the magnetization remained out-of-plane at remanence across the wire due to the much higher PMA. The 24 nm thick center (lower PMA) started to reverse at $H_z = -22$ kOe, while the 16 nm and the 8 nm thick regions (higher PMA) reversed at $H_z = -24$ kOe and $-28$ kOe, respectively. The modest changes in reversal field reflect the relatively small variation of PMA with thickness, which is qualitatively consistent with the experimental observations. We also see that the simulated reversal are much higher than experiments because the direction of $K_u$ was the same for all cells leading to coherent reversal at much higher fields than in experiments.

In conclusion, a double-layer resist with an undercut profile produced liftoff features with reduced edge crowning and a tapered thickness profile. The tapering leads to a modulation in PMA at the edges: in [Co (0.6 nm)/Pd (1.2 nm)]$_{20}$ wires the PMA was higher at the center while in $L1_0$-FePt grown on MgO (001) substrate the PMA was higher at the edges. This opposite thickness-dependence of PMA is related to the origin of anisotropy in both systems, and affects whether the edge or center of the wire reverses at lower field. The differences are more prominent in [Co/Pd]$_{20}$. The ability to tune PMA via the edge profile provides a tool for controlling the reversal process and could also affect other edge-related properties such as ferromagnetic resonant modes [16] or domain wall [17] or skyrmion dynamics.

**Acknowledgements**

The authors thank Prof. Jingsheng Chen from National University of Singapore for providing the facilities to fabricate [Co/Pd]$_n$ multilayer samples. The authors also gratefully acknowledge support from C-SPIN, a STARnet Center of SRC supported by DARPA and MARCO, and NSF award ECCS1101798. Shared experimental facilities from the NanoStructures Laboratory and the Center for Materials Science and Engineering under NSF award DMR1419807 were used.

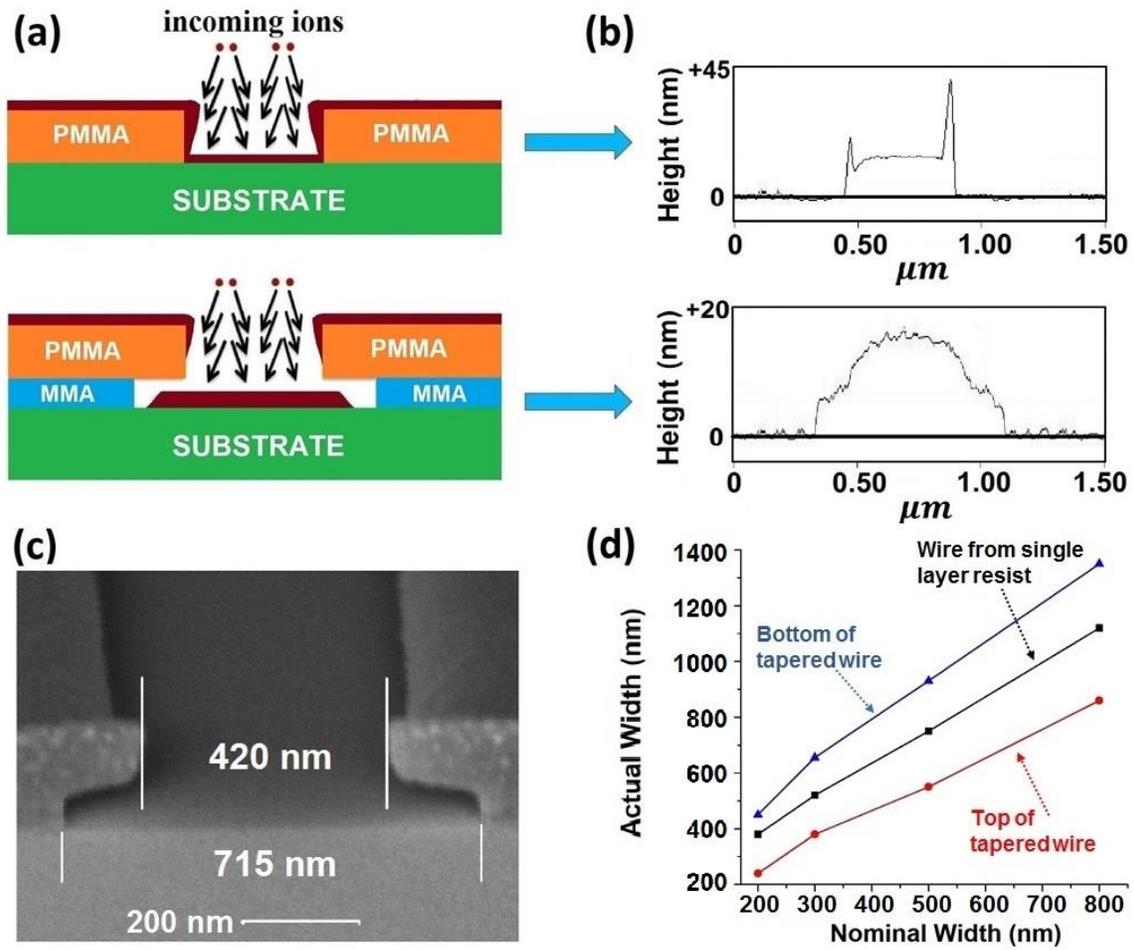

**Figure 1.** (a) Schematics of liftoff using single-layer resist and a double-layer resist with undercut profile; (b) AFM profiles of nanowires patterned by sputtering and liftoff from single-layer resist and double-layer resist, the latter has a tapered profile; (c) SEM of exemplary double-layer resist; (d) Plot of actual wire width vs. nominal wire width for samples using single-layer resist and double-layer resist.

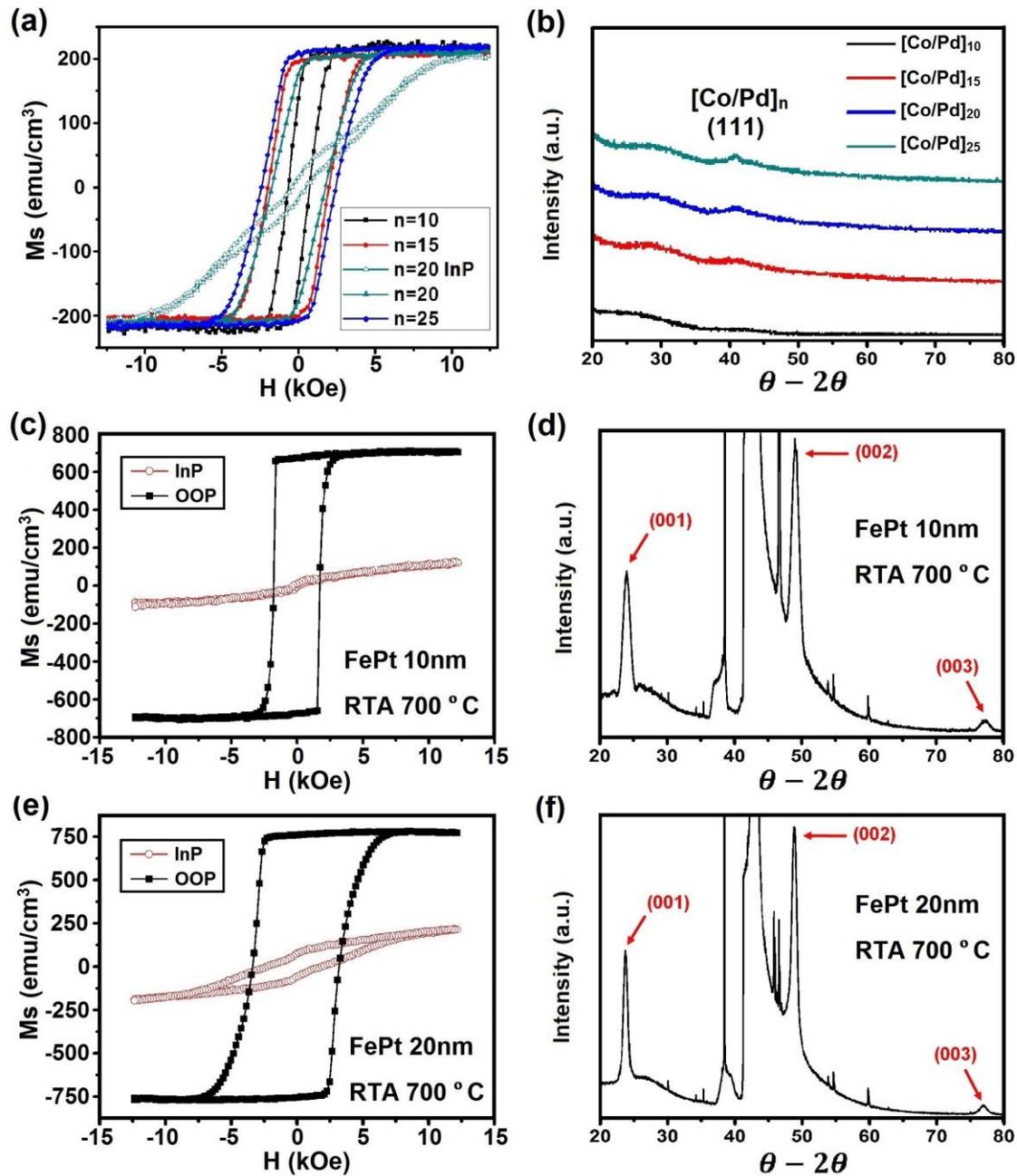

**Figure 2.** (a) Out-of-plane hysteresis loop of continuous [Co(0.6 nm)/Pd (1.2 nm)]$_n$ films with different repetitions *n*. Both in-plane and out-of-plane loops for n=20 are included. (b) XRD showing the (111) peak in [Co/Pd]$_n$ arising from the textured film growth. (c,e) Out-of-plane and in-plane hysteresis loop of annealed continuous $L1_0$-FePt films grown on MgO (001) substrate with thickness of (c) 10 nm and (e) 20 nm; (d,f) XRD of (c,e), respectively, indicating $L1_0$-ordered crystalline structure. (111) for $L1_0$-FePt ($\approx 41.1°$) is covered by the MgO main peak and cannot be observed.

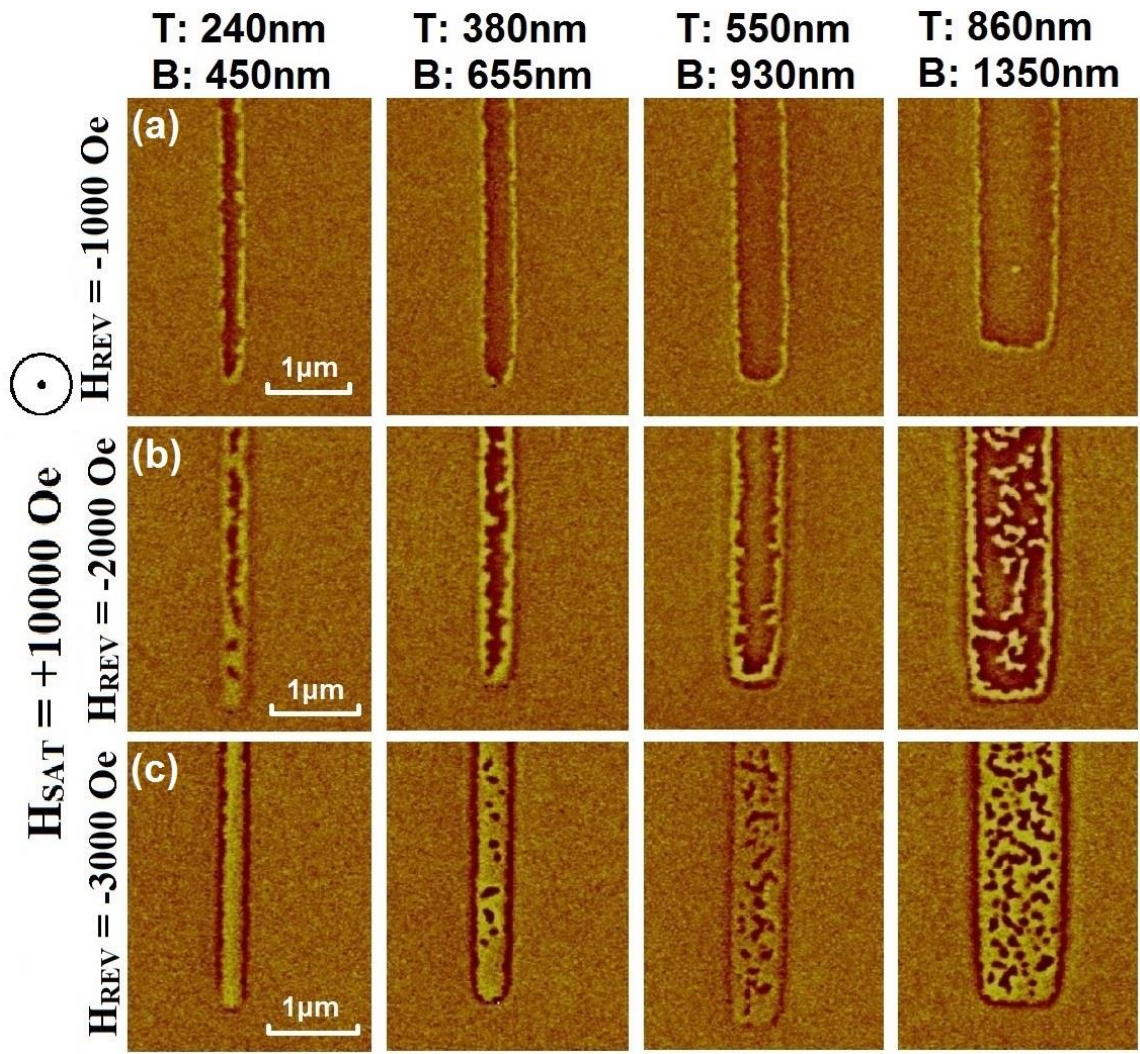

**Figure 3**. MFM images of [Co/Pd]$_{20}$ nanowires prepared with double-layer resist, samples were first saturated with an out-of-plane field of H$_{SAT}$ = +10 kOe then a reverse field of H$_{REV}$ was applied and samples imaged at remanence. (a) H$_{REV}$ = -1000 Oe; (b) H$_{REV}$ = -2000 Oe; (c) H$_{REV}$ = -3000 Oe. T and B represent the width at the top and base. Reverse domains appear bright.

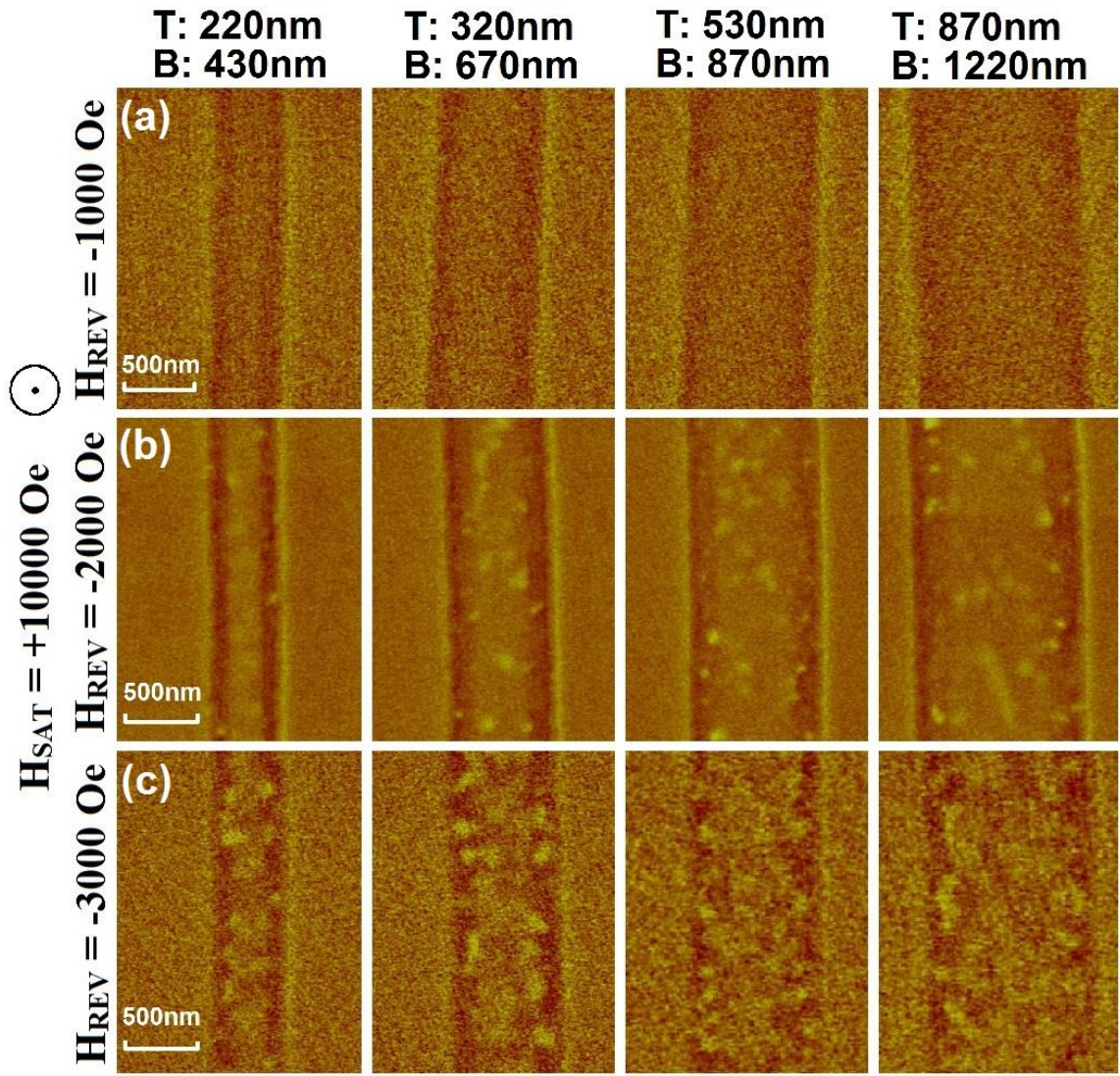

**Figure 4.** MFM images of $L1_0$-FePt nanowires prepared with double-layer resist, samples were first saturated with an out-of-plane field of $H_{SAT}$ = +10 kOe then a reverse field of $H_{REV}$ was applied and samples imaged at remanence. (a) $H_{REV}$ = -1000 Oe; (b) $H_{REV}$ = -2000 Oe; (c) $H_{REV}$ = -3000 Oe. T and B represent the width at the top and base. Reverse domains appear bright.

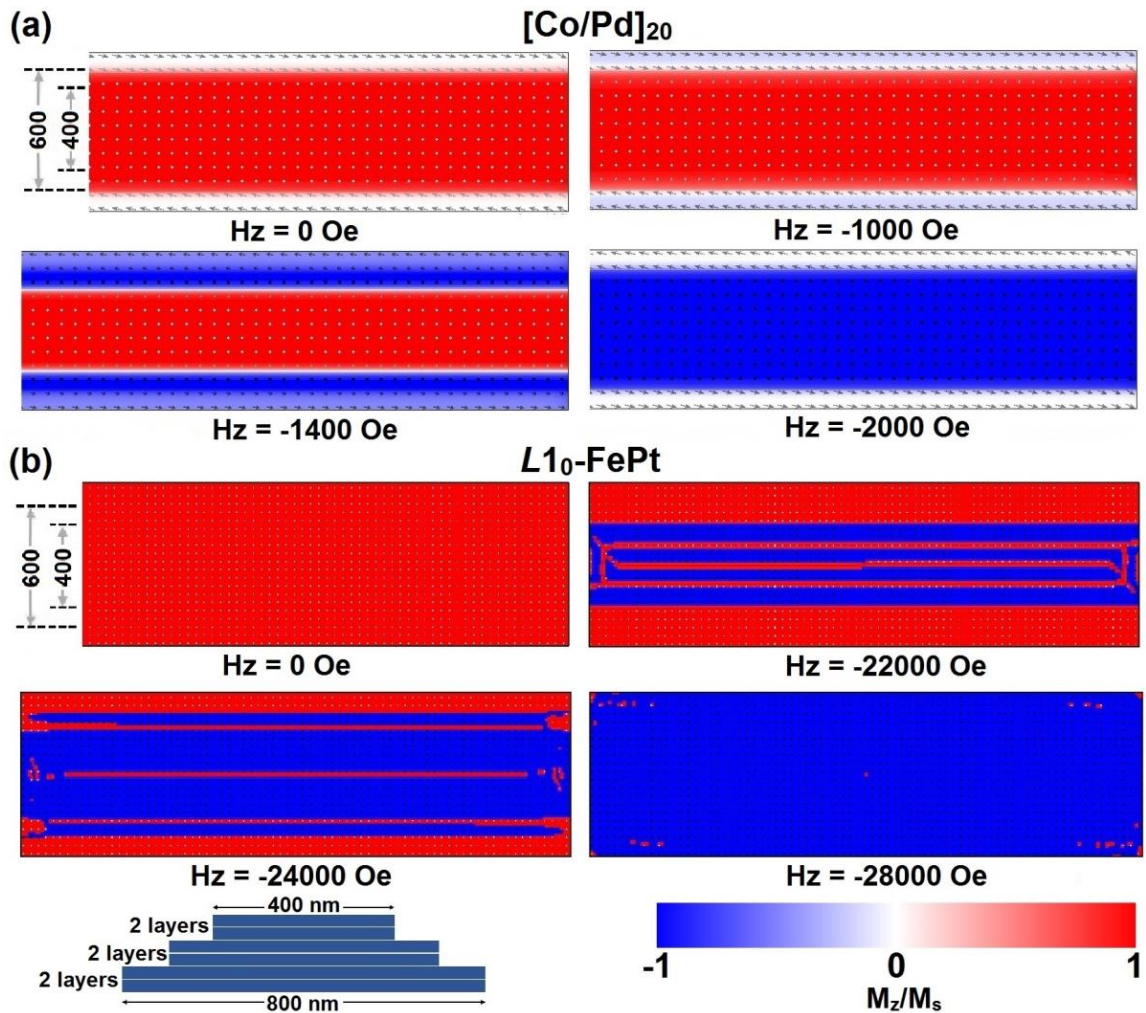

**Figure 5.** *OOMMF* simulations of [Co/Pd]$_{20}$ and $L1_0$-FePt nanowires with PMA, samples are first saturated in the +z (out-of-plane) direction. The images show only the widest layer at the base. The positions of the middle layer (600 nm) and the top layer (400 nm) of cells are labeled in the first panel. Red and blue shading indicate the direction of M$_z$ (up or down). The arrows show the projection of the magnetization vector (white or black dots represent out-of-plane arrows). The images show the magnetization state of (a) [Co/Pd]$_{20}$ nanowires measured in the presence of field H$_z$; (b) $L1_0$-FePt nanowires measured at remanence after applying field H$_z$.